\documentclass[twocolumn,showpacs,superscriptaddress,preprintnumbers,amsmath,amssymb]{revtex4}
\usepackage{dcolumn}
\usepackage{bm}
\usepackage{verbatim}
\usepackage{graphicx}
\usepackage[english]{babel}
\newcommand{\kb}{k_{\scriptsize B}}

\newcommand{\init}{\rm in}
\newcommand{\out}{\rm out}

\newcommand{\z}[1]{#1}
\newcommand{\y}[1]{#1}
\newcommand{\x}[1]{#1}
\begin{document}

\title{Reduced coherence in double-slit diffraction of neutrons}
\author{R. Tumulka}\email{tumulka@everest.mathematik.uni-tuebingen.de}\affiliation{Mathematisches
Institut, Eberhard-Karls-Universit\"at, Auf der Morgenstelle 10,
72076 T\"ubingen, Germany}
 \author{A. Viale}\email{viale@ge.infn.it} \author{N.
  Zangh\`\i}\email{zanghi@ge.infn.it}\affiliation{Dipartimento di
  Fisica, Istituto Nazionale di Fisica Nucleare, Sezione di Genova,
  Via Dodecaneso 33, 16146 Genova, Italy}

\date{\today}

\begin{abstract}
In diffraction experiments with particle beams, several effects lead
to a fringe visibility reduction of the interference pattern. We
theoretically describe the intensity one can measure in a double-slit
setup and compare the results with the experimental data obtained
with cold neutrons. Our conclusion is that for cold neutrons the
fringe visibility reduction is due not to decoherence, but to initial
incoherence.
\end{abstract}


\maketitle

%%%%%%%%%%%%%%%%%%%%%%%%%%%%%%%%%%%%%%%%%%%%%%%%%%%%%%%%%%%%%%%%%%%%%%%%%%%%%%%%%%%%%
%%%%%%%%%%%%%%%%%%%%%%%%%%%%%%%%%%%%%%%%%%%%%%%%%%%%%%%%%%%%%%%%%%%%%%%%%%%%%%%%%%%%%
\section{Introduction}
\begin{figure}
\begin{center}
\includegraphics [scale=0.80]{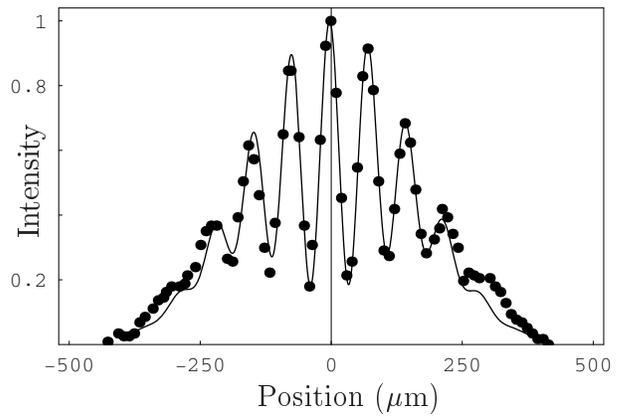}
\caption{Comparison between theoretical prediction derived here
(solid line) and experimental data taken from~\cite{zeil} (solid circles) 
for neutron double-slit diffraction.}\label{fig:fit}
\end{center}
\end{figure}

%%%%%%%%%%%%%%%%%%%%%%%%%%%%%%%%%%%%%%%%%%%%%%%%%%%%%%%%%%%%%%%%%%%%%%%%%%%%%%%%%%%%%
%%%%%%%%%%%%%%%%%%%%%%%%%%%%%%%%%%%%%%%%%%%%%%%%%%%%%%%%%%%%%%%%%%%%%%%%%%%%%%%%%%%%%
We provide a theoretical description of the intensity pattern in
double-slit experiments with neutrons, with specific attention to the
\emph{cold} neutron diffraction ($\lambda~\approx~20\,\mathrm{\AA}$)
carried out by Zeilinger \emph{et al}.~in 1988~\cite{zeil}. The result 
we obtain is shown in Fig.~\ref{fig:fit}.

Usually, the main problem in the analysis of diffraction experiments
is to establish exactly which causes bring about the \y{reduced}
coherence experimentally inferred from the detected signal. This effect
can be produced both by the initial preparation of the beam (namely,
the non-dynamical \emph{incoherence}) and by the interaction of its
constituting particles with the environment (namely, the dynamical
\emph{decoherence})~\cite{pfau, alten, wise, giulini, alicki,
pascazio, alcim, horn, angel, horn2}. 

We find that in the experiment of Ref.~\cite{zeil} decoherence
does not play any role in the fringe
visibility reduction, which indeed is entirely due to incoherence of
the source. This \z{conclusion} is opposite to that of
Ref.~\cite{angel}, where it is claimed that decoherence is essential
for explaining the data from~\cite{zeil}.

We provide calculations and numerical simulations in support of an unexpected
incoherence cause, which we propose plays a role in explaining
the experimental data of Ref.~\cite{zeil}. This cause is that the
width of the wave function impinging on the double-slit is
comparable with the size of the double-slit setup 
(i.e., the distance between the two slits and the slit apertures). 
We argue that this feature of the incoming wave function leads to a 
slight difference in transverse momentum between the
two wave packets emanating from the grating, such that the
centers of the packets move apart. This momentum difference
was already suggested in Ref.~\cite{angel}, but no physical
explanation was given. It proves relevant for fitting the
data of Ref.~\cite{zeil}. Even though it does not reduce
the fringe visibility, it can significantly affect the shape of the
interference pattern, changing the overlap of the two
packets and so the position of secondary minima and maxima.
%
%
%
%%%%%%%%%%%%%%%%%%%%%%%%%%%%%%%%%%%%%%%%%%%%%%%%%%%%%%%%%%%%%%%%%%%%%%%%%%%%%%%%%%%%%
%%%%%%%%%%%%%%%%%%%%%%%%%%%%%%%%%%%%%%%%%%%%%%%%%%%%%%%%%%%%%%%%%%%%%%%%%%%%%%%%%%%%%
\section{Visibility reduction in the intensity pattern}
%%%%%%%%%%%%%%%%%%%%%%%%%%%%%%%%%%%%%%%%%%%%%%%%%%%%%%%%%%%%%%%%%%%%%%%%%%%%%%%%%%%%%
%%%%%%%%%%%%%%%%%%%%%%%%%%%%%%%%%%%%%%%%%%%%%%%%%%%%%%%%%%%%%%%%%%%%%%%%%%%%%%%%%%%%%
%
%
We now describe the various \y{causes of reduced coherence} which
can occur within neutron interferometry. \y{Let the $y$ direction be the direction of 
propagation of the beam. We assume that the grating is 
translation invariant in the $z$ direction, and that the motion in the $y$ direction 
is essentially classical, so that the problem reduces to one dimension, corresponding 
to the $x$ axis.}
%%%%%%%%%%%%%%%%%%%%%%%%%%%%%%%%%%%%%%%%%%%%%%%%%%%%%%%%%%%%%%%%%%%%%%%%%%%%%%%%%%%%%
\subsection{Initial incoherence\label{subsec:incoh}}
%%%%%%%%%%%%%%%%%%%%%%%%%%%%%%%%%%%%%%%%%%%%%%%%%%%%%%%%%%%%%%%%%%%%%%%%%%%%%%%%%%%%%
In the initial preparation of a beam, it is problematic to keep a
perfect control on the \emph{monochromaticity} and on the
\emph{collimation} of the beam.

\y{Concerning the non-monochromaticity,} every
spectral component of the beam contributes incoherently to \y{every}
other. In \z{our case, assuming} that the detection screen \z{is}
\y{parallel to the $x$ axis,}
the intensity \z{observed} on the screen \z{at coordinate $x$} is
\begin{equation}\label{eq:velocity}
I (x) = \int \mathrm{d}\lambda \: f(\lambda) \: I_\lambda(x),
\end{equation}
where $f(\lambda)\mathrm{d}\lambda$ is the wavelength distribution of
the beam and $I_\lambda(x)$ the intensity corresponding to a single
wavelength $\lambda$. Thus, one can continue the analysis with a
single wavelength and postpone the integration \eqref{eq:velocity} to
the final step (see Sec.~\ref{sec:measured}).

Concerning the collimation of the beam, \y{a relevant cause of incoherence
is} that the particle
source emits \emph{random wave functions}, whose randomness lies in
the deviation of its \z{direction of propagation from the $y$
direction, corresponding to imperfect collimation.} \y{This cause
of incoherence will be taken
into account in our model of the beam by means of a random wave vector $k$.}
%%%%%%%%%%%%%%%%%%%%%%%%%%%%%%%%%%%%%%%%%%%%%%%%%%%%%%%%%%%%%%%%%%%%%%%%%%%%%%%%%%%%%
\subsection{Decoherence}\label{sec:decoherence}
%%%%%%%%%%%%%%%%%%%%%%%%%%%%%%%%%%%%%%%%%%%%%%%%%%%%%%%%%%%%%%%%%%%%%%%%%%%%%%%%%%%%%
In the treatment of decoherence, we make explicit use of the fact
that we are concerned with neutrons, in particular at the low
energies of Ref.~\cite{zeil}. In this case, in fact, the only
relevant decoherence channel---if any---consists in collisions with
air molecules, so that the dynamics can be modelled within a
Markovian description of the scattering event, in particular in the
\emph{large scale} approximation allowed by air molecules
(see~\cite{gallis,giulini,alcim}). The model \z{obtained in this way
allows an estimate} of the neutron \emph{coherence time} $\tau_{\rm
coh}$~\cite{gallis, alcim}:
\begin{equation}\label{taucoh}
\tau_{\rm coh} = \frac{1}{P(\Theta_{\mathcal{E}})\,\sigma_{\rm
tot}}\:\sqrt{\frac{8}{\pi\,\kb\, \Theta_{\mathcal{E}}\, m_{\rm
air}}}\:,
\end{equation}
where $P(\Theta_\mathcal{E})$ is the environmental pressure at the
temperature $\Theta_\mathcal{E}$, $\sigma_{\rm tot}$ the total cross
section of the scattering events, $\kb$ the Boltzmann constant and
$\:m_{\rm air}\approx 4.8\cdot 10^{-26}\,\mathrm{Kg}$ the
mean mass of air molecules. \z{(Eq.~\eqref{taucoh} takes into account
a correction by a factor $2\pi$, usually missing in the literature,
that was theoretically predicted in~\cite{sipe, vacchini} and
experimentally checked in~\cite{horn}.)}

The result is that $\tau_{\mathrm{coh}}$ is much greater than the
time of flight, even in extreme experimental situations. In fact,
even considering a surrounding pressure of $1\,\rm atm$ (though
typically ``the beam paths along the optical bench" are ``evacuated
in order to minimize absorption and scattering"~\cite{zeil}) and room
temperature, for an estimated total cross section of
$10^{-27}\,\mathrm{m^2}$ \footnote{To our knowledge, direct values of
$\sigma_{\rm tot}$ under the considered conditions for the 
molecules which compose the air  are not reported in the literature.
Nevertheless, starting from related measurements~\cite{cross} and
using typical trends of the cross section, it is possible to
estimate very roughly the bound $\sigma_{\rm tot} \lesssim
10^{-27}\,\mathrm{m^2}$. However, even if a correct estimation of
$\sigma_{\rm tot}$ led to a greater value, $\tau_{\rm coh}$ 
would still be much larger than the time of flight $T$ since
five orders of magnitude lie between our estimate for 
$\tau_{\rm coh}$ and $T$. Furthermore, we have assumed very unfavorable
experimental circumstances.} we obtain that $\tau_{\rm coh} \approx 140\,\mathrm{s}$.
This time---not much smaller than the neutron lifetime---shows that
the coherence is fully kept for the duration of most experiments. For
instance, in~\cite{zeil} the mean time of flight is $T \approx
0.023\,\mathrm{s}$, several orders of magnitude smaller than
$\tau_{\mathrm{coh}}$.

For this reason, in the following we shall regard the neutron beam
as uncoupled from its environment and shall describe the dynamics
through the usual unitary Schr\"odinger evolution.

\z{This conclusion contradicts that of Sanz
\emph{et al.}\ in Ref.~\cite{angel}, who claim that
``decoherence is likely to exist in Zeilinger \emph{et al.}'s
experiment." Indeed, their own calculations do not support their
conclusion. The basis of their claim is that the
\emph{damping term} $\Lambda_t$ in their expression for
the observed intensity pattern [see their Eq.~(27)] turns out,
when fitted to the data, to be nonzero. However, this term
could just as well represent \emph{incoherence} instead of decoherence
(as does for example the similar quantity $\mathcal{A}$ in their Eq.~(16),
or the \emph{coherence length} in Eq.~(47) of Ref.~\cite{alcim})
\footnote{In particular, the caption of Fig.~2 in \cite{angel} is incorrect in so far
as Fig.~2(a) does not include incoherence, and Fig.~2(b) may represent
either incoherence or decoherence.}. Indeed, our estimate of the coherence time shows that
the damping $\Lambda_t$ cannot be attributed to decoherence.}
%%%%%%%%%%%%%%%%%%%%%%%%%%%%%%%%%%%%%%%%%%%%%%%%%%%%%%%%%%%%%%%%%%%%%%%%%%%%%%%%%%%%
%%%%%%%%%%%%%%%%%%%%%%%%%%%%%%%%%%%%%%%%%%%%%%%%%%%%%%%%%%%%%%%%%%%%%%%%%%%%%%%%%%%%%
\section{Model for the intensity pattern\label{sec:measured}}
%%%%%%%%%%%%%%%%%%%%%%%%%%%%%%%%%%%%%%%%%%%%%%%%%%%%%%%%%%%%%%%%%%%%%%%%%%%%%%%%%%%%%
%%%%%%%%%%%%%%%%%%%%%%%%%%%%%%%%%%%%%%%%%%%%%%%%%%%%%%%%%%%%%%%%%%%%%%%%%%%%%%%%%%%%%
In order to obtain a theoretical description of the intensity pattern measured in the 
experiment, we precisely describe the whole evolution of the neutron beam, from 
its production to the arrival on the detection screen.

\y{We set up a concrete model for what the wave functions of the
neutrons in the beam look like, and thus obtain
the density matrix \x{and the intensity pattern}.
(Another approach \cite{englert} is to guess the density matrix from
information theoretic principles, but we prefer to avoid the
invocation of such principles.)}

The width $A$ of the 
\emph{entrance} slit fixes the spatial extension of the (random) wave packets. 
The wave packets $\psi_0$ produced by the source are modeled, when passing the entrance
slit, as Gaussian wave packets with mean $0$ and standard deviation $s_0 = A/\sqrt{12}$, i.e.,
the standard deviation of the \emph{uniform} probability distribution over the interval of length
$A$. Thus, \[\psi_0(x) \sim \exp \Bigl(
-\frac{x^2}{2s_0^2} + ikx\Bigr) , \] where we take the random wave number
$k$ to have a Gaussian distribution with mean 0 and standard deviation
$\sigma_k$.  As established in Sec.~\ref{sec:decoherence}, away from the grating, the neutrons follow 
the free Schr\"odinger evolution, so that---after the flight over the distance $L_0$---the packet 
just before the grating (the \emph{incoming} wave function) has the Gaussian form
\begin{equation}\nonumber
\psi_{\init}(x) \sim \exp\bigg\{-(1-i\gamma)\frac{\big[x-\delta(k)\big]^2}{2s^2}+ ikx\bigg\},
\end{equation}
with $\gamma = \lambda L_0/(4\pi s_0^2)$, $s=s_0\sqrt{1+\gamma^2}$ and $\delta(k)=L_0 \lambda k/(2\pi)$.
In Sec.~\ref{sec:model} we shall present two different models for 
the wave function immediately after the grating (the \emph{outgoing} wave function $\psi_{\out}$) 
as a function of the incoming wave function $\psi_{\init}$. 

We now consider the free propagation of the outgoing state $\psi_{\out}$ toward the screen. 
We obtain the density matrix $\rho_\mathrm{out}$ by averaging
$|\psi_\mathrm{out}\rangle \langle \psi_\mathrm{out}|$ with respect to the
probability distribution of the wave number $k$. The intensity measured on
the \emph{distant} screen is proportional to the diagonal elements of the
density matrix $\rho_T$ at the \emph{time of arrival} $T = m L \lambda
/(2\pi\hbar)$ (with $m$ the mass of the
neutron)---as already stated and well motivated in the
literature~\cite{flux, alcim}: $I_\lambda(x) \sim \rho_T(x,x)$. 
Before identifying it with the measured intensity,
we have also to consider the integration~\eqref{eq:velocity} over the wavelength and 
the finite spatial resolution $x_0$ of the detector. For the latter, we shall consider \emph{flat response} 
on the interval $[x\!-\!x_0/2\,,\,x\!+\!x_0/2]$ around each position $x$, 
so that the intensity is finally given by
\begin{equation}\nonumber
I (x) = \frac{1}{x_0}\int\! \mathrm{d}\lambda \: f(\lambda)
\!\int_{x-x_0/2}^{x+x_0/2}
\!\!\!\!\!\!\!\!\!\!\!\mathrm{d} y \:\: I_\lambda(y),
\end{equation}
where $x_0$ and $f(\lambda)$ are directly adoptable from Ref.~\cite{zeil}.
%%%%%%%%%%%%%%%%%%%%%%%%%%%%%%%%%%%%%%%%%%%%%%%%%%%%%%%%%%%%%%%%%%%%%%%%%%%%%%%%%%%%%
\section{Passage through the grating\label{sec:model}}
%%%%%%%%%%%%%%%%%%%%%%%%%%%%%%%%%%%%%%%%%%%%%%%%%%%%%%%%%%%%%%%%%%%%%%%%%%%%%%%%%%%%%
We describe two models for the passage of the neutron
through the grating. The first is a simplified model, which can be treated
analytically and features the momentum difference between the two
transmitted packets in the transverse $x$ direction. The second, on which 
Fig.~\ref{fig:fit} is based, is more flexible and achieves better agreement with
experimental data.
%%%%%%%%%%%%%%%%%%%%%%%%%%%%%%%%%%%%%%%%%%%%%%%%%%%%%%%%%%%%%%%%%%%%%%%%%%%%%%%%%%%%%
\subsection{First model\label{subsec:simplified}}
%%%%%%%%%%%%%%%%%%%%%%%%%%%%%%%%%%%%%%%%%%%%%%%%%%%%%%%%%%%%%%%%%%%%%%%%%%%%%%%%%%%%%
The central assumption of this model is that the outgoing wave function is of the form
\begin{equation}\nonumber
\psi_{\out}(x)=F(x) \psi_{\init}(x),
\end{equation}
where $F$ is a complex function whose modulus is $\leq 1$
and signifies the degree of transmission, while the phase signifies the
phase shift during the passage through the grating. The simplest model
assumption would be to set $F(x) = 1$ for $x \in [x_1-a_1/2,x_1+a_1/2]
\cup [x_2-a_2/2, x_2 + a_2/2]$ and $F(x) =0$ otherwise, where the two
intervals represent the two slits, their apertures $a_1$, $a_2$ and
distance $d$ are known \cite{zeil}, and $x_j = (-1)^j (a_j +d)/2$ for $j=1,2$
is the center of the $j$-th slit. With this choice of $F$, the
multiplication by $F$ is a projection operator. However, to enable
analytical computations we find it more useful to set
\begin{equation}\nonumber
F(x) \sim \frac{1}{\sigma_1} \exp\bigg[-\frac{(x-x_1)^2}{2\sigma_1^2}\Big]
+ \frac{1}{\sigma_2} \exp\Big[-\frac{(x-x_2)^2}{2\sigma_2^2}\bigg],
\end{equation}
where $\sigma_j= a_j/\sqrt{12}$.

\begin{figure}
\begin{center}
\includegraphics [scale=0.80]{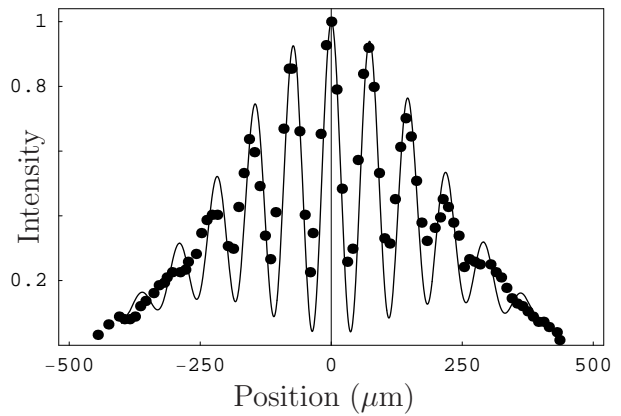}
\caption{Comparison between the theoretical prediction derived with our
\emph{first model} for the passage through the grating
(solid line) and experimental data taken from~\cite{zeil} (solid circles).}\label{fig:fit2}
\end{center}
\end{figure}

Now $I_\lambda(x)$ can be calculated exactly. In Fig.~\ref{fig:fit2} the comparison between
the theoretical intensity which follows from this model and the experimental data of
Ref.~\cite{zeil} is shown. Apart from a constant background subtracted from experimental data,
there is only one fit parameter $\sigma_k$, for which we have found $\sigma_k = 10^5\,\mathrm{m/s}$.
This value corresponds to a $x-$velocity with standard deviation equal to
$0.006\,\mathrm{m/s}$, while the mean $y-$velocity distribution is $214\,\mathrm{m/s}$~\cite{zeil}.

It is useful to look at the explicit formula for the outgoing
wave function. For simplicity, set $k=0$. Then
\begin{equation}\nonumber
\psi_{\out}(x) = \psi_{{\out}, 1}(x) + \psi_{{\out}, 2}(x),
\end{equation}
where 
\begin{equation}\nonumber
\psi_{{\out}, j}(x) \sim \exp \Big[ -(x-\xi_j)^2 (\alpha_j - \frac{i\gamma}{2s^2}) + 
\frac{i p_j x}{\hbar}\Big]
\end{equation}
and where we have introduced the following quantities, for $j=1,2$: $\alpha_j=1/(2s^2)+1/(2\sigma_j^2)$, 
$\xi_j = x_j/(2 \alpha_j \sigma_j^2)$ and  $p_j = \hbar \gamma
x_j/ (\sigma_j^2 + s^2)$. 
Therefore $\psi_{\out}$ is the sum of two Gaussian packets with momentum in the 
$x$ direction equal to $p_1$ and $p_2$. 
Moreover, since $x_1$ and $x_2$ have opposite signs, the same holds for $\xi_1$ and $\xi_2$ and, consequently, 
for $p_1$ and $p_2$. In other words, the packets produced by the grating have two opposite momenta in the transverse direction, as a consequence of the finite width of the neutron wave function 
compared to the scale fixed by the grating size. (Note that if $\psi_{\init}$ is very spread out, i.e.,
if $s$ is large, then $p_j \sim (\sigma_j^2 + s^2)^{-1}$ is small).

Also in the presence of a common drift expressed by $k$, the packets produced by the 
grating have wave number in the transverse direction equal to $k+p_1/\hbar$ and $k+p_2/\hbar$, 
with analytical expressions for the wave functions and $p_j$ more complicated than before. 

Even though the agreement with experimental data is not completely satisfactory, this model
illustrates the origin of the transverse momenta $p_1$ and $p_2$ that, as 
already noticed by Sanz \emph{et al.} in Ref.~\cite{angel}, move outwards the
$x-$position of secondary minima.

\y{The values $p_j$} can also be predicted numerically by solving the Schr\"odinger
equation for the passage of the neutron through the grating, \x{treating
the problem as two-dimensional with the double-slit modelled as an impenetrable potential barrier}.
\y{Our numerical simulations were not accurate enough to yield \x{realistic} values, \x{but}
did confirm that the $p_j$ are effectively nonzero, point in opposite directions (outwards),
and have roughly equal modulus.} Numerical simulations also suggest
 that the values of $p_j$ are substantially the same for
all $\lambda$ within the range of wave lengths selected for the experiment.
%%%%%%%%%%%%%%%%%%%%%%%%%%%%%%%%%%%%%%%%%%%%%%%%%%%%%%%%%%%%%%%%%%%%%%%%%%%%%%%%%%%%%
\subsection{Second model\label{subsec:refined}}
%%%%%%%%%%%%%%%%%%%%%%%%%%%%%%%%%%%%%%%%%%%%%%%%%%%%%%%%%%%%%%%%%%%%%%%%%%%%%%%%%%%%%
The second model allows to fit $p_1$ and $p_2$
to the data. It is based on the assumption that the outgoing wave function is of the form
\begin{equation}\nonumber
  \psi_\mathrm{out}(x)  = c_1 \, \varphi_1(x)
  + c_2 \, \varphi_2(x)\,,
\end{equation}
where $c_j \, \varphi_j$ is the wave packet emanating from the
left ($j=1$) or right ($j=2$) slit, which we assume to be of Gaussian
form:
\begin{equation}\nonumber
  \varphi_j(x) = \frac{1}{(2\pi \sigma_j^2)^{1/4}} e^{-(x-
  x_j)^2/4\sigma_j^2} \, e^{ik_j x}
\end{equation}
allowing for the possibility that the drift $k_j$ and the weight $c_j$ are not equal for the
two packets, where $\sigma_j$ and $x_j$ are the same quantities as introduced in
Sec.~\ref{subsec:simplified}. 

We write the wave numbers $k_j$ as $k_j = k +p_j/\hbar$, i.e., the sum of the wave number $k$ 
of the incoming packet and the transverse momentum $p_j$ acquired during the passage through the grating, 
as previously highlighted.

We estimate the coefficients $c_j$ as
\begin{equation}\nonumber 
c_j = c_j(k) = \big\vert \psi_{\init}(x_j)\big\vert \,.
\end{equation}
{Thus the wave packet emanating from slit 1 has greater weight $|c_1|^2$ if
the incoming wave packet is closer to slit 1. (For further discussion
see Ref.~\cite{tesi}.)}

Fig.~\ref{fig:fit} shows the resulting prediction of this model. 
The following parameters were obtained by means of a
\emph{fit procedure}: the two velocities $p_1/m = -0.0034\,\mathrm{m/s}$ and $p_2/m =
0.0029\,\mathrm{m/s}$ \footnote{\z{These values of $p_1$ and $p_2$
agree qualitatively with the estimates in Ref.~\cite{angel}.
Nonetheless, the estimation in~\cite{angel} is dubious, as it is based on an
incomprehensible application of the Heisenberg uncertainty
principle.}}, and the angular divergence of the \x{beam
$\sigma_{k}=4976\,\mathrm{m}^{-1}$ (which corresponds to a $x-$velocity distribution with
standard deviation equal to $0.0003\,\mathrm{m/s}$)}. 
Moreover, \x{also in this case} a constant background has been subtracted from experimental data. 
The $p_j$ have been assumed to be the same for all wave lenghts $\lambda$ selected for the experiment.
%%%%%%%%%%%%%%%%%%%%%%%%%%%%%%%%%%%%%%%%%%%%%%%%%%%%%%%%%%%%%%%%%%%%%%%%%%%%%%%%%%%%%%
%%%%%%%%%%%%%%%%%%%%%%%%%%%%%%%%%%%%%%%%%%%%%%%%%%%%%%%%%%%%%%%%%%%%%%%%%%%%%%%%%%%%%

\end{document}